\documentclass[pra,aps]{revtex4}
\usepackage {graphicx}

\begin{document}
\title{THEORETICAL AND EXPERIMENTAL APPROACH TO SPIN DYNAMICS IN THIN 
MAGNETIC FILMS.}

\author{C. Tannous, A. Fessant, J. Gieraltowski, J.P. Jay, D. Spenato,\\
J. Langlois\dag, and H. Le Gall\P}

\affiliation{Université de Bretagne Occidentale, Laboratoire d'Etude des Matériaux,
\dag Laboratoire des Collisions Electroniques et Atomiques,\\
6, Avenue le Gorgeu - BP: 809 - 29285 Brest Cedex, FRANCE,\\
\P CNRS, LMIMS, 1, place Aristide Briand - 92195 Meudon Cedex, FRANCE}

\date{December 4, 2000}

\begin{abstract}
The Landau-Lifshitz (L-L) equation describing the time 
dependence of the magnetisation vector is numerically integrated fully 
without any simplifying assumptions in the time domain and the magnetisation 
time series obtained is Fourier transformed (FFT) to yield the permeability 
spectrum up to 10 GHz. The non linear results are compared to the 
experimental results obtained on magnetic amorphous thin films of Co-Zr, 
Co-Zr-Re. We analyse our results with the frequency response obtained 
directly from the Landau-Lifshitz equation as well as with the second order 
Gilbert frequency response.

\end{abstract}

\pacs{PACS numbers: 75.70.Pa, 75.60.Ch, 75.70.Cn, 75.70.-i} 

\maketitle

\section{INTRODUCTION}

The development of high-density information storage involves developments in 
new materials for recording heads and recording media. The required 
properties for recording heads are high saturation magnetisation 
$ {M_{s}}$, low coercivity $ {H_{c}}$, low magnetostriction coefficient 
$\lambda $ and high permeability $\mu$ over a broad frequency 
range (typically several hundred MHz). \\

Thin films ~\cite{spenato} such as Co-M1-M2, (M1, M2 = Zr, Nb, Re...), 
FeTaN and FeZrN ~\cite{vanderiet} have been investigated in view of 
these applications. These materials belong to a new class of ultrasoft films 
used in high density magnetic recording devices. In order to improve the 
data read-out in magnetic recording heads the frequency behaviour of the 
permeability should stay large at high frequencies. At low frequencies, when 
the exciting field is applied along the easy axis, the initial permeability, 
due mainly to motion of magnetic walls, is large and rolls off at about a 
few tens of kHz. On the other hand, when the exciting field is applied 
perpendicular to the easy axis, the permeability is generally lower and due 
to spin rotation. In this case, the roll-off frequency may approach a few 
hundreds MHz.\\

Previous studies agree with the fact the Landau-Lifshitz 
equation describes successfully the ferromagnetic behaviour of the material 
from the resonance frequency and linewidth points of view.\\

Since the Landau-Lifshitz equation has been used in the study of many 
magnetic systems and more recently in thin magnetic films ~\cite{spenato}, 
we applied it to Co-Zr, Co-Zr-Re thin amorphous films. 

We perform our studies experimentaly in the [10 MHz, 10 GHz] frequency range 
with a special broadband measurement technique ~\cite{fessant} that we have 
developed previously.\\

Usually, the Landau-Lifshitz equation, is linearised and Fourier transformed 
in order to extract the complex magnetic permeability as a function of 
frequency. Instead, we take the fully non linear equation and treat it with 
time domain integration methods.\\

We develop a new accurate and stable numerical time integration technique 
over long durations (in order to have accurate results in the frequency 
domain) and perform the Fourier transform (FFT) directly on the calculated 
magnetisation time series. These results are compared with the measurement 
and agree to some extent with the previous theoretical results obtained with 
the linearisation approach. Nevertheless non linear effects are visible in 
the frequency range we consider.\\

This paper is organised as follows. In section 2 we present our experimental 
and simulation results and discuss the effects of some physical parameters 
on the behaviour of the permeability. The experimental results are compared 
to the linear and non linear responses and section 3 contains our 
conclusions.\\

\section{EXPERIMENTAL AND SIMULATION RESULTS}

The Landau-Lifshitz equation is:

\begin{equation}
 {\frac{d\textbf{M}}{dt}} =  { \gamma (\textbf{M}\times \textbf{H})
-\alpha \gamma \frac { \textbf{M} \times (\textbf{M} \times \textbf{H} ) }{ \mid \textbf{M}\mid } } 
\end{equation}

where $\textbf{M}$ is the magnetisation vector, \textbf{H} the effective field, 
$\gamma $ the gyromagnetic ratio (in our case $\gamma =2.2.10^{5} mA^{-1}.s^{-1}$) and $\alpha $ 
the damping parameter.\\
 
The effective field is:

\begin{equation}
 { \textbf{H}= H_{k} \textbf{x} + h_{y} cos(\omega t) \textbf{y}-M_{z} \textbf{z}  }
\end{equation}

where $ {H_k}$ is the anisotropy field and $ {M_z}$ is the 
demagnetisation field.

The frequency permeability response $\mu (\omega) = \mu^{\prime} (\omega) -j \mu^{\prime\prime}(\omega)$ 
 might be calculated directly from the L-L equation or after a 
perturbation expansion to the second order in $\alpha $ as done previously 
by Gilbert ~\cite{spenato,vanderiet}. 

The direct (small amplitude) Landau-Lifshitz frequency response is given by: 

\begin{equation}
\mu (\omega) = 1  + \frac{ A + j\omega B}  {( \omega _{0}^{2} - \omega^{2} ) + j \omega D}
\end{equation}

with:

\begin{equation}
 { A = \gamma ^{2} M_{s}^{2} (1 + \alpha ^{2} )( 1 + \frac{ H_{k} }{ M_{s} } )  } 
\end{equation}
and:

\begin{equation}
 { B = \alpha \gamma M_{s}}
\end{equation}

Moreover:

\begin{equation}
 { C = \omega _{0}^{2} = \gamma ^{2}H_{k}M_{s}(1 + \alpha ^{2})(1 + \frac{H_k}{M_s})  }  
\end{equation}

and,
\begin{equation}
 { D = \alpha \gamma M_{s}( 1 + 2\frac{H_k}{M_s})}
\end{equation}

The reason behind this is the assumption that $\alpha $ is small (smaller 
than 0.1) as confirmed experimentally.\\

The Gilbert and the L-L responses are compared in Fig. 1. As one expects, 
good agreement is visible between the two responses for small values of the 
damping parameter $\alpha $. For larger values of $\alpha $ (typically 
larger than 0.5) the responses start to differ and this difference would 
grow larger when non linear effects start to come into play.\\

In order to calculate the magnetisation as a function of time, we employ a 
fast time integration technique based on an accurate fourth order 
Runge-Kutta time domain integration scheme. The L-L equation is solved in 
the rotating frame (Larmor precession pulsation $\omega $ = $\gamma $ 
$ {H_k}$) and the system is excited with a Dirac pulse in time. The FFT 
is then performed to get the frequency response. Since the amplitude of the 
exciting field ($ {h_y}$) is small, the previous results might be 
directly compared with the linear response.\\

The studied $Co_{100-x}Zr_{x}$, and $Co_{100-x}ZrRe_x$ ($3 < x< 14$) amorphous thin films 
(of about 1 $\mu $m thickness) were deposited on 25 mm diameter glass 
substrate using conventional diode-sputtering equipment. The amorphous state 
has been verified by electron probe microanalysis and the thickness obtained 
by a profilometer. The materials present an in-plane magnetisation. and 
their saturation magnetisation $ {M_s}$, measured on a vibrating sample 
magnetometer, is higher than $10^{6} A.m^{-1}$. The anisotropy field 
$ {H_k} $  has been investigated with a B-H loopmeter and from 
ferrromagnetic resonance linewidth.\\

A new broad-band method for the determination of complex magnetic 
permeability, described in ~\cite{spenato}, has been used.\\ 

The theoretical results are compared to experimental ones obtained on two 
kind of samples. The first one  $Co_{95}Zr_5$ has a high uniaxial anisotropy 
field ($ {H_k}=2200 A.m^{-1}$) and a well defined anisotropy axis. The second 
one ($Co_{95}Zr_5$) has a smaller anisotropy field ($ {H_k}= 400 A.m^{-1}$) and a 
distribution of the anisotropy axis. Further details of sample preparation 
are given in ~\cite{spenato}.\\

The first example of experimental spectra of the complex permeability is 
shown in figures 2 and -3. Good agreement between theoretical and 
experimental values can be observed. The linear and the non linear responses 
are compared to the experimental results. Real and imaginary parts are 
displayed separately in figure 2 for the L-L case and figure 3 for the non 
linear case.\\

The second example displayed in figures 4-5 concerns a  $Co_{95}Zr_5$ thin 
amorphous film. The agreement between experiment and calculation is not so 
good in this case as already explained in ~\cite{spenato} where the experimental data 
were compared to the Gilbert frequency response. The discrepancy might be 
attributed to the distribution of the anisotropy orientation. The real and 
imaginary parts of the full non linear response are displayed in figure 5.\\

In the set of figures we vary the value of $ {\alpha}$ at will in 
order to estimate the interplay of linear and non linear effects. The 
results show that in spite of the relatively large variations in 
$ {\alpha}$ , the non linear response is close to the linear response 
in all cases. Nevertheless, the skewness of the imaginary part of the 
permeability follows better the non linear results. \\

In all cases, linear and non linear results for the real part of permeability obey the 
Stoner-Wohlfarth limit at small frequencies ~\cite{spenato}. Non linear effects are 
visible but not strong enough to produce large departures from the linear 
results because of the small values of $\alpha $ considered.

\section{CONCLUSION}

This work considers the complex permeability at high frequency of thin 
ferromagnetic films, when the magnetisation is due to spin rotation. We have 
calculated, using linear and non-linear Landau-Lifshitz theory, the 
frequency dependence of the permeability of a uniaxial thin film with an 
in-plane anisotropy. We have shown the influence of non-linearity through 
the damping term $\alpha $ on the permeability spectra.\\

In the first batch of experimental results, good agreement with the model 
using Landau-Lifshitz theory is obtained whereas in the second batch of 
experimental measurements, we do not observe such a good agreement with 
theory.\\

Previously, other works considered eddy currents and the dispersion of 
uniaxial anisotropy as potential sources of discrepancy between theory and 
experiment. In this work, we investigated the effects of the non-linearity 
and confirm that it is not an issue in these materials in spite of an 
improvement in the skewness of the imaginary part of the permeability.

\vfill
\centerline{\Large\bf Figure Captions}

\begin{itemize}
\item[Fig.\ 1:]Comparison between the Gilbert and the direct
L-L frequency responses ($\alpha $= 0.03, $ {H_{k}} = 2200 A.m^{-1}$ 
and $ {M_{s}}=1.15.10^{6} A.m^{-1}$.

\item[Fig.\ 2:]Real and imaginary part of the direct L-L 
frequency response and the  measured spectrum of an amorphous 
$Co_{95}ZrRe_{5}$ thin film when the exciting field is perpendicular
to the easy  axis. Thickness is 1 $\mu $m, $ {H_{k}} = 2200 A.m^{-1}$ and 
$ {M_{s}}=1.15.10^{6} A.m^{-1}$.

\item[Fig.\ 3:]Real and imaginary part of the simulated non linear response and 
measured complex permeability frequency 
spectrum of an amorphous $Co_{95}ZrRe_{5}$ thin film when the exciting field is 
perpendicular to the easy axis. Thickness is 1 $\mu $m, $ { H_{k}} = 2200 
A.m^{-1}$ and $ { M_{s}}=1.15 .10^{6} A.m^{-1}$.

\item[Fig.\ 4:]Real and imaginary parts of  the  direct 
L-L linear response and the  measured frequency spectrum of an 
amorphous $Co_{95}Zr_5$ thin film when the exciting field is perpendicular
 to the easy axis. Thickness is 1 $\mu$m, $ {H_{k}} = 400 A.m^{-1}$ and 
$ {M_{s}}=1.2.10^{6} A.m^{-1}$.

\item[Fig.\ 5:]Real and imaginary parts of the simulated non linear response and the  measured complex 
permeability frequency spectrum of an amorphous  $Co_{95}Zr_5$ thin film when the 
exciting field is perpendicular to the easy axis. Thickness is 1 $\mu $m, 
$ {H_{k}}=400 A.m^{-1}$ and $ {M_{s}}=1.2. 10^{6} A.m^{-1}$.

\end{itemize}

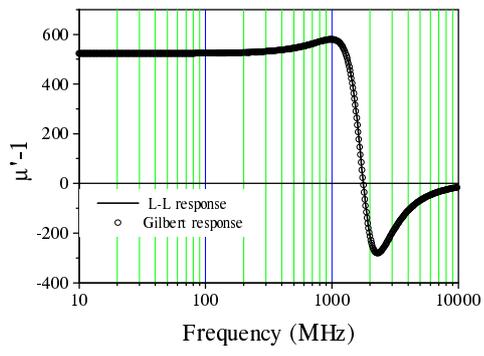
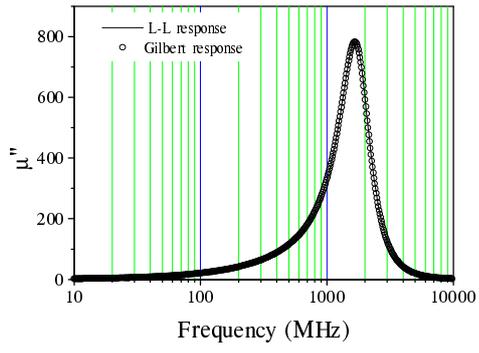
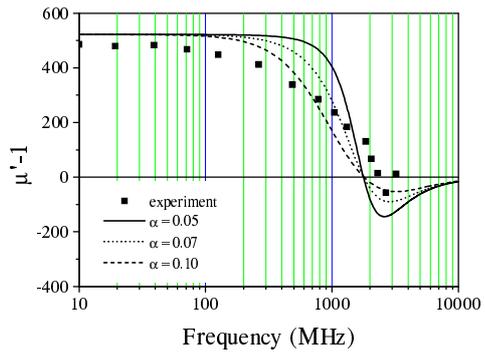
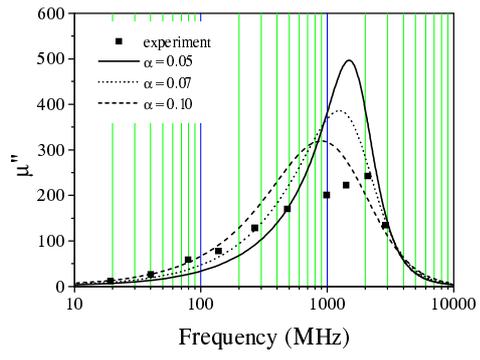
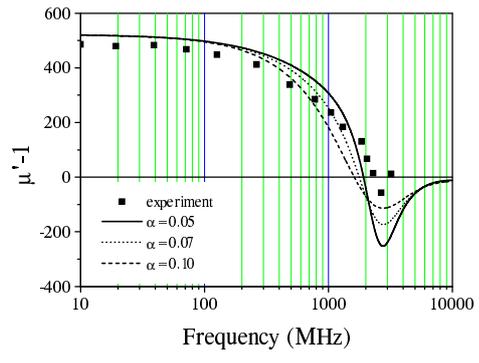
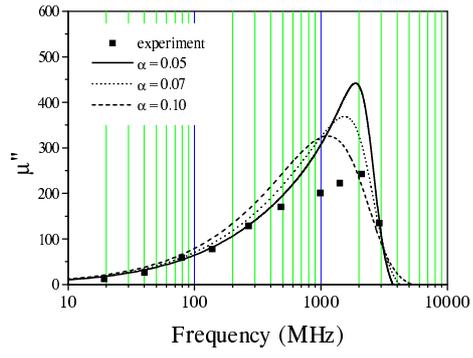
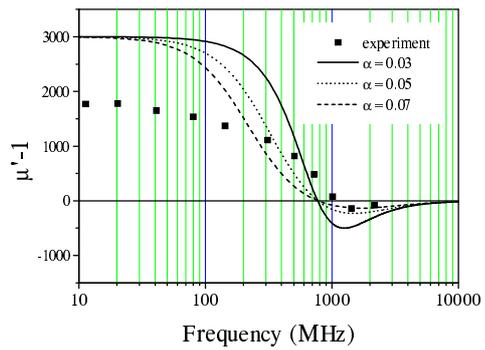
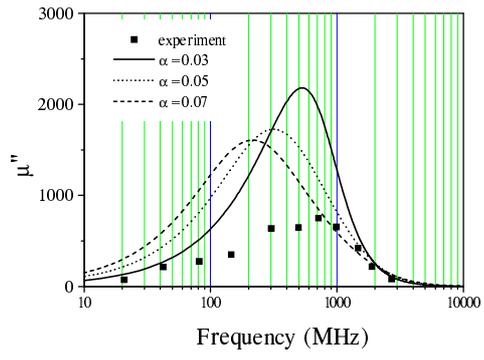
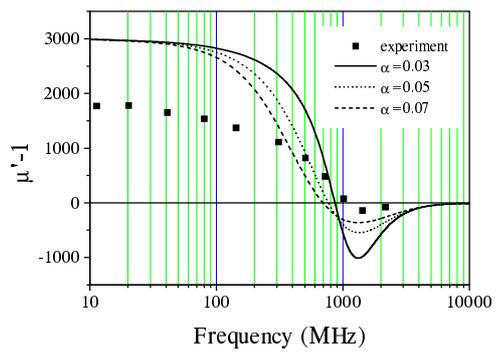
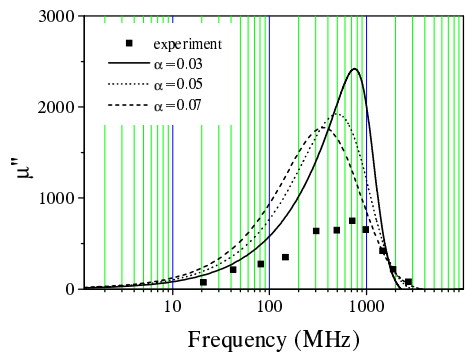
\end{document}